\documentclass[%
reprint,
 amsmath,amssymb,
 aps,showkeys,
prl
]{revtex4-1}

\usepackage{graphics}
\usepackage{graphicx}
\usepackage{dcolumn}
\usepackage{bm}
\usepackage{hyperref}
\usepackage{placeins}
\usepackage{xcolor}
\usepackage[english]{babel}

\begin{document}



\title{Universal performance bounds of restart}

\author{Dmitry Starkov$^{1,2}$ and Sergey Belan$^{1,3}$}
\email{sergb27@yandex.ru}
\affiliation{$^{1}$Landau Institute for Theoretical Physics, Russian Academy of Sciences, 1-A Akademika Semenova av., 142432 Chernogolovka, Russia}
\affiliation{$^{2}$National Research University Higher School of Economics, Faculty of Mathematics, Usacheva 6, 119048 Moscow, Russia}
\affiliation{$^{3}$National Research University Higher School of Economics, Faculty of Physics, Myasnitskaya 20, 101000 Moscow, Russia}

\begin{abstract} 
As has long been known to computer scientists, the performance of probabilistic algorithms characterized by  relatively large runtime fluctuations can be improved by applying a restart, i.e., episodic interruption of a randomized computational procedure followed by initialization of its new statistically independent realization.
A similar effect of restart-induced process acceleration   could potentially be possible in the context of enzymatic reactions, where  dissociation of the enzyme-substrate intermediate corresponds to restarting the catalytic step of the reaction. 
To date, a significant number of analytical results have been obtained in physics and computer science regarding the effect of restart on the completion time statistics in various model problems, however,  the fundamental limits of restart efficiency remain unknown. 
Here we derive a range of  universal 
statistical inequalities that offer constraints on  the  effect that restart could impose on the  completion time of a generic stochastic process.
The corresponding bounds are expressed via simple statistical metrics of the original process such as harmonic mean $h$, median value $m$ and mode $M$, and, thus, are remarkably practical. 
We test our analytical predictions with multiple numerical examples, discuss  implications arising from them and important avenues of future work.
\end{abstract}

\maketitle

\vskip \baselineskip

\textit{Introduction.} 
Restarting was first proposed as a promising optimization tool of probabilistic algorithms in the early 1990s \citep{Alt_1991,Luby_1993}.
The restart-induced speed up may seem counterintuitive at first glance, but the basic idea behind this technique is quite transparent: if the current realization of the randomized algorithm takes too long, it may be faster (on average) to retry attempt with a
new random seed to avoid prolonged wandering in the region of the configuration space far from the actual solution. Since then, restarting has become a routine procedure used to hasten computational  tasks whose run-time exhibits significant fluctuations \cite{Lorenz_2018,Lorenz_2019,Lorenz_2021,Montanari_2002, Moorsel_2004, Wu_2006,Gagliolo_2007,Wu_2007,Streeter_2007,Wedge_2008,Roulet_2017,Gomes_1997, Biere_2009, Huang_2007,Wolter_thesis}.
In particular, the option of restart is built into state-of-art constraint satisfaction problem solvers  \cite{Schulte_2010,Cire_2014,Amadini_2018, Wallace_2020} (see also \cite{Schroeder_2001,Moorsel_2006} for other computer science applications).

A current wave of interest to this topic from the statistical physics community has been sparked by the work
of Evans and Majumdar \cite{EM_2011}  who showed that stochastic (Poisson) restart expedites diffusion-mediated search.
After that it has been demonstrated on a number of  different examples that a specially selected restart frequency makes it possible to minimize the average time for completing random search tasks \cite{Evans_2011,Whitehouse_2013,Evans_2014,Kusmierz_2014,Kusmierz_2015,Pal_2016,Eule_2016,Nagar_2016,Kusmierz_2019,Ray_2020,Singh_2020,Ahmad_2020,Radice_2021,Faisant_2021,Calvert_2021,Mercado_2021,Bonomo_2021,Tucci_2022,Ahmad_2022, Chen_2022,Abdoli_2021,Ray_2019a,Santra_2020a,Evans_2016c,Evans_review_2020,Chechkin_2018b}.
Also, recent development of a model-independent renewal approach, originally proposed for the purposes of describing the single enzyme kinetics \cite{Reuveni_2014}, has provided a shortcut to the exact completion time statistics of an arbitrary stochastic process under an arbitrary restart protocol \cite{Rotbart_2015,Reuveni_PRL_2016,Reuveni_PRL_2017}.
This  fruitful approach helped to reveal unexpected  universality in statistics of  optimally restarted processes \cite{Reuveni_PRL_2016}, to establish remarkably simple sufficient conditions for when restart is beneficial \cite{Reuveni_PRL_2016,Pal_JPA_2022,Eliazar_JPA_2021} and to rigorously quantify  impact of restart on various statistical characteristics of random processes \cite{Belan_PRL_2018, Belan_PRR_2020,Eliazar_arxiv_2022,Eliazar_arxiv_2022b}.



The natural course of development of the research field  poses  the following question to us:
What, if any, are the fundamental limitations  of the optimization via restart?
The knowledge of the exact completion time distribution allows to determine the optimal restart strategy (which may be to not restart at all) and the  corresponding best possible performance for a given stochastic process.
In practice, however, the complete statistics of the completion time is usually unavailable, see, e.g., \cite{Luby_1993,Gagliolo_2007,Reuveni_2014,Lorenz_2018,Lorenz_2021,Wu_2007,Streeter_2007}.
The existing literature lacks understanding of how to evaluate the potential performance of restart based  on some limited set of statistical characteristics of an observable process.
In this Letter we fill this gap by exploring how much restart can lower the expected completion time of a stochastic process with partially specified properties.
More specifically, we present the universal lower bound  on the mean completion time  respected by any stochastic process under an arbitrary restart protocol.
Besides, we construct the universal upper bound on the mean completion time of generic stochastic process at optimal restart conditions. 
Both types of probabilistic inequalities are generalized to the high order statistical moments of random completion time.
Being formulated in terms of easily interpreted statistical characteristics of an unperturbed stochastic process, the resulting bounds   provide valuable insight into what  restart can and cannot do.
As a useful corollary, our analysis provides a novel sufficient condition for restart to be beneficial which requires very little  information on the statistics of the random process of interest. 
Finally, we propose a broad generalization of the well-known inequality offering constraint on  the relative fluctuation of the expected completion time of optimally restarted processes.

\textit{Model formulation.}
Consider 
a generic stochastic process which ends after a random time $T$ if allowed to take place without
interruptions.
Statistical properties of the variable $T$ are described by the probability density $P(T)$ with a proper normalization $\int_0^{\infty}dTP(T)=1$. 
This implies that the process terminates in finite time with probability 1.
As discussed below, our key results remain unchanged or require a trivial  modification when one introduces the non-zero probability of never stopping.

The restart protocol ${\cal R}$ is characterised by a (possibly infinite)  sequence of   inter-restarts time intervals $\tau_1,\tau_2,\dots$.
If the process is completed prior to the first restart event, the story ends there.
Otherwise, the process will start from scratch and begin  anew. Next, the process may either complete prior to the second restart or not, with the same rules.
This procedure repeats until the process finally reaches completion.
Importantly, we assume that  protocol ${\cal R}$ is  uncoupled from the process internal dynamics: the restart decisions do not use information on the current internal state of the process.

In the simplest case of strictly periodic protocol, which is of particular methodological importance as explained below,  the process is restarted whenever $\tau$ units of time  pass.
The  expected value of the random  completion  time $T_\tau$ of the process subject to such a restart procedure can be obtained by averaging of an appropriate renewal equation.
The result is given the following expression 
\begin{equation}
\label{completion_time}
\langle T_{\tau}\rangle = \frac{\int_0^{\tau} P(T)TdT+\tau\int_{\tau}^\infty P(T)dT}{\int_0^{\tau} P(T)dT},
\end{equation}
which, thus, relates the expectation of the random completion time in the presence of periodic restart to the statistics of the "bare" (i.e. restart-free) process.

\textit{Restart performance is limited to a quarter of the harmonic mean completion time.}
First of all, 
  we  seek to derive inequality of the form $\langle T_{\cal R}\rangle\ge C{\cal T}$, where $T_{\cal R}$ denotes the random completion time of the generic stochastic process under arbitrary restart protocol ${\cal R}$, ${\cal T}$  is expressed through some simple statistical characteristics   of the  original process (such as  statistical moments, quantiles or mode of the  probability density $P(T)$), 
 and $C$ is the universal positive constant which depends neither on specific form of $P(T)$ nor on the particular restart schedule ${\cal R}$.

Previous works have shown the importance of relative fluctuation $\sigma/\mu$, where $\mu=\langle T\rangle$ and $\sigma=\sqrt{\langle T^2\rangle-\langle T\rangle^2}$, for the analysis of the potential response of stochastic process to restart.
Namely, the inequality $\sigma/\mu>1$ represents a sufficient condition for the existence of a restart protocol that reduces the expected completion time \cite{Rotbart_2015,Reuveni_PRL_2016,Pal_JPA_2022}.
Given this result, let us first find out if knowledge of the mean value $\mu$ and the standard deviation $\sigma$ allows one to write a lower bound on the average performance of  restart.
Consider, probability density $P(T)=p\delta(T-t_1)+(1-p)\delta(T-t_2)$, where $0\le t_1\le t_2$ and $0\le p\le 1$.
Putting  $t_2= \frac{\mu^2+ \sigma^2}{\mu}$, $p=\frac{\sigma^2}{\mu^2 +\sigma^2}$, $\tau=t_1+0$ and $t_1\to 0$,
one immediately obtains From Eq. (\ref{completion_time}) $\langle T_\tau\rangle=t_1/p \to0$.
We see that for the fixed  values of $\mu$ and $\sigma$, the completion time $\langle T_\tau\rangle$ can be arbitrarily small.
Therefore, the pair  $(\mu$,$\sigma)$ does not produce any non-trivial lower bound.

Our derivation of the desired lower bound limit is based on the special properties of the periodic restart strategy.
As shown by Luby et al. \cite{Luby_1993}  for discrete time case and generalized to continuous settings by Lorenz \cite{Lorenz_2021} (see also \cite{SM} for   simpler and even more general proof),
if you found a value $\tau_{\ast}\ge0$ (probably $\tau_{\ast}=+\infty$) such that $\langle T_{\tau_{\ast}}\rangle\le \langle T_{\tau}\rangle$ for any $\tau\ge 0$, then $\langle T_{\tau_{\ast}}\rangle\le \langle T_{\cal R}\rangle$ for all ${\cal R}$. 
In other words, optimally tuned periodic restart beats any other restart strategy.
In addition, the same authors have proved (see also \cite{SM})  that   the mean performance of an optimal periodic restart obeys the condition
\begin{equation}
\label{inequality_from_Luby}
\langle T_{\tau_\ast} \rangle\ge \frac{1}{4} \min_{\tau} \frac{\tau}{Pr[T\le\tau]}.
\end{equation}

\begin{figure}
\centering
\includegraphics[scale=0.32]{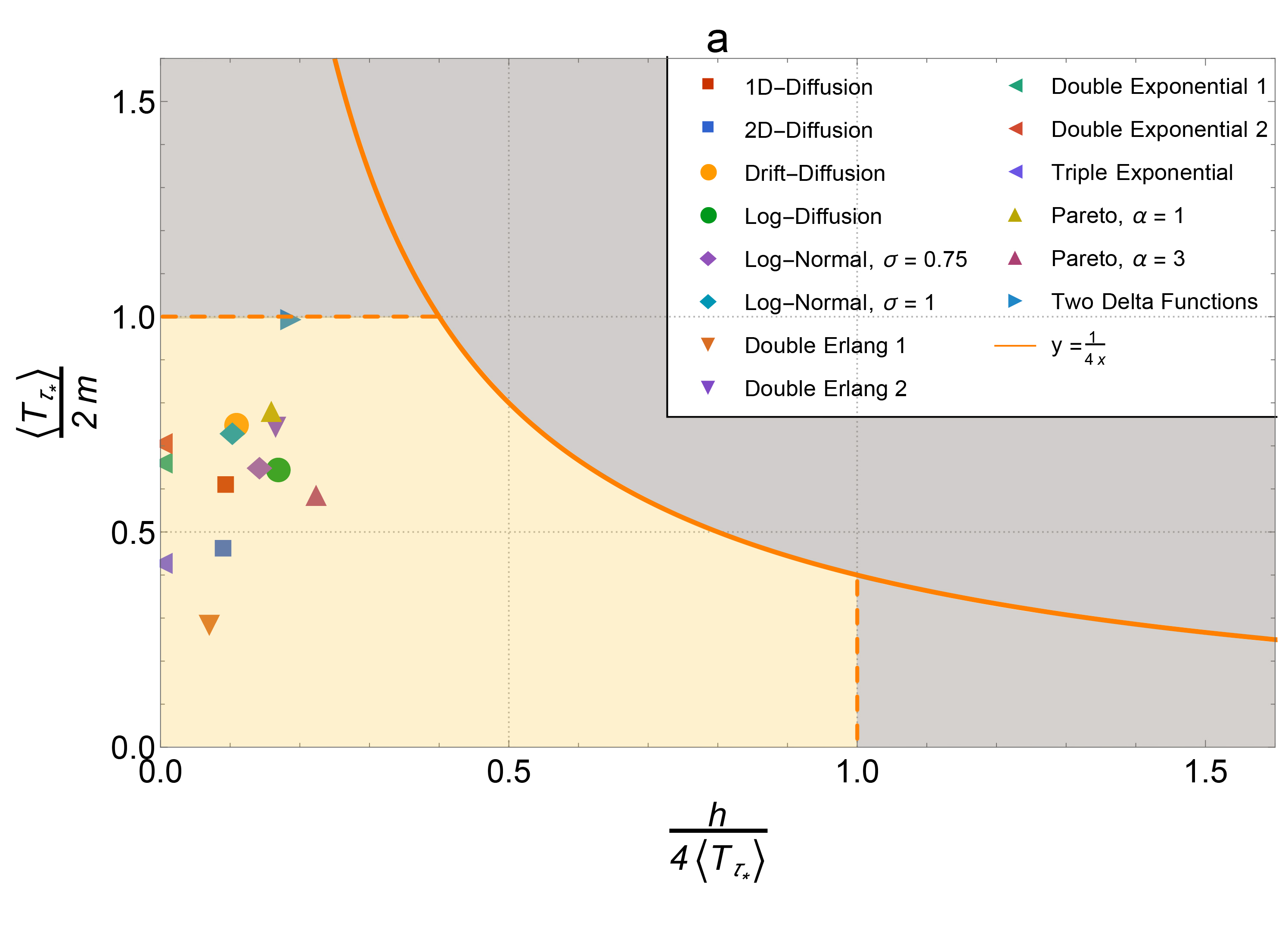}
\label{fig:4}
\caption{ Diagram in the $x-y$ plane, where $x=h/(4\langle T_{\tau_\ast}\rangle)$ and  $y=\langle T_{\tau_\ast}\rangle/(2m)$,  depicting relationship between mean completion time of an optimally restarted process $\langle T_{\tau_\ast}\rangle$, the median  $m$ and the harmonic mean $h$ of the original process.
As indicated in a legend,
each point in these diagrams  corresponds to a specified completion time distribution $P(T)$. 
In accordance with Eqs. (\ref{our_third_result}) and (\ref{our_first_result})  and probabilistic  inequality $h\le2m$ 
all points  belong to the light orange region determined by the conditions $0\le x\le 1$, $0\le y\le 1$, $y\le 1/(4x)$.}
\label{fig:4}
\end{figure}

Let us show that using Eq. (\ref{inequality_from_Luby}) together with the optimal property of the periodic restart strategy leads to a simple  performance bound of restart.
Namely, applying   Markov's  inequality \cite{Gallager_2013} to the variable $\omega=1/T$ we find $\text{Pr}[T\le \tau]=\text{Pr}[\omega\ge\frac{1}{\tau}]\le \tau\langle \omega\rangle =\tau \langle \frac{1}{T}\rangle$.
Next, taking into account Eq. (\ref{inequality_from_Luby}) one  obtains  $\langle T_{\tau_\ast}\rangle\ge \frac14 h$, where $h=\langle T^{-1}\rangle^{-1}$  is the harmonic mean completion time of the original process.
And finally, since $\langle T_{\cal R}\rangle\ge \langle T_{\tau_{\ast}}\rangle$ for any ${\cal R}$, this   yields 
\begin{equation}
\label{our_third_result}
\langle T_{\cal R}\rangle\ge \frac14 h.
\end{equation}
No constraints have been imposed on the form of  $P(T)$,  and, therefore, Eq. (\ref{our_third_result})  is universally valid for any setting. 
What is more, this estimate remains valid also when stochastic process may have non-zero probability of never ending, if $h$ is always understood as the harmonic mean completion time of the halting trials \cite{note0}.


\textit{Particular case of smooth unimodal distribution.}
Somewhat less general, but still informative,  result can be obtained  if we assume that  the completion time distribution $P(T)$ is smooth and exhibits single local maximum at some non-zero value of $T$. 
This class of probability densities covers, in particular, vast number of random search models, see, e.g., Refs. \cite{EM_2011,Evans_2014,Evans_review_2020,Kusmierz_2014,Kusmierz_2015,Singh_2020,Abdoli_2021,Tucci_2022,Ray_2019a,Santra_2020a,Evans_2016c}.
The efficiency of any restart protocol in this case satisfies the inequality
\begin{equation}
\label{our_forth_result}
\langle T_{{\cal R}}\rangle\ge \frac14 M,
\end{equation}
where $M=\text{argmax}_TP(T)>0$ is the mode of the probability distribution $P(T)$, i.e. the value of the random completion time $T$ that occurs most frequently. 
To prove this result let us introduce $\tau_0\equiv\text{argmin}_\tau\frac{\tau}{\text{Pr}[T\le \tau]}$.
Clearly, assumption  $M>0$ implies that  $\tau_0>0$.
Since the smooth function  $f(\tau)=\frac{\tau}{\text{Pr}[T\le \tau]}$ attains its minimal value at $\tau=\tau_0$, one obtains $df(\tau_0)/d\tau=0$ or, equivalently, $P(\tau_0)\tau_0=\int_0^{\tau_0} P(T)dT$.
Next, as the unimodal function $P(T)$ is non-decreasing on the interval form $0$ to $M$,
this extrema condition implies the inequality $\tau_0\ge M$ and, therefore, $\frac{\tau_0}{\text{Pr}[T\le \tau_0]}\ge \frac{M}{\text{Pr}[T\le \tau_0]}\ge M$. 
Together with Eq. (\ref{inequality_from_Luby}) this yields inequality $\langle T_{\tau_\ast}\rangle\ge \frac14 M$. 
Recalling that $\langle T_{\cal R}\rangle\ge\langle T_{\tau_\ast}\rangle$ for all  ${\cal R}$, we then obtain 
Eq. (\ref{our_forth_result}).
Note also, that if the  probability distribution $P(T)$ has  multiple local maxima, then  $\langle T_{{\cal R}}\rangle\ge \frac14 M^{min}$, where $M^{min}$ is the leftmost mode. 
Moreover, similarly to Eq. (\ref{our_third_result}), inequality (\ref{our_forth_result}) holds even for potentially non-stopping processes, with the obvious caveat that $M$ should now be considered as the most frequent completion time of halting trials.


\textit{ The twice median sets upper bound on the optimized mean completion time.}
Having considered the lower bounds, we now turn to the  opposite question.
How good is the best restart strategy?
In other words, we wish to construct an
inequality of the form $\langle T_{\tau_\ast}\rangle\le C{\cal T}$, where the time scale ${\cal T}$  is determined by  the  properties of the original stochastic process, and $C$ is the universal positive constant which depends  neither on specific form of $P(T)$ nor on the optimal restart period $\tau_\ast$.


It is easy to understand that the upper bound limit on optimal performance cannot be expressed via the harmonic mean $h$ or the mode $M$. 
Indeed, for  the half-normal distribution $P(T)=\sqrt{\frac{2}{\pi\sigma^2}}e^{-\frac{T^2}{2\sigma^2}}$ one has  $\tau_\ast=+\infty$, so that $\langle T_{\tau_\ast}\rangle=\langle T\rangle>0$, whereas $h=M=0$.
Therefore, inequalities of the form $\langle T_{\tau_\ast}\rangle\le C_1 h$ and $\langle T_{\tau_\ast}\rangle\le C_2M$, where $C_1$ and $C_2$ are positive constants, cannot be universally valid.


The desired universal upper bound can be expressed in terms of the median completion time $m$ of the original process obeying by definition the equation $Pr[T\le m]=1/2$.
Indeed, taking into account that $\langle T_{\tau_\ast}\rangle\le \langle T_{\tau}\rangle$ for any $\tau\ge0$ together with the  inequality $\langle T_{\tau}\rangle\le  \frac{\tau}{Pr[T\le \tau]}$, which straightforwardly follows from  Eq. (\ref{completion_time}), we find  $\langle T_{\tau_\ast}\rangle\le  \frac{\tau}{Pr[T\le \tau]}$.
Substituting  $m$ for  $\tau$ in the last inequality one obtains 
\begin{equation}
\label{our_first_result}
\langle T_{\tau_\ast}\rangle\le 2m.
\end{equation}
Thus, no matter how heavy the tails of $P(T)$ are, in the presence of an optimally tuned periodic restart, the average completion time does not exceed twice the median of the unperturbed process. 
More generally, if the process has non-zero probability of never halting, we arrive at the estimate 
 $\langle T_{\tau_\ast}\rangle\le 2m_s/q$, where  $m_s$ denotes the   median completion times of the halting trials, whereas  $q$ is the probability that process ends for a finite time \cite{SM}.

Importantly, the bound dictated by Eq. (\ref{our_first_result}) is sharp.  
Indeed, for $P(T) = \frac{1}{2} \delta(T - t) + \frac{1}{2}\delta(T - 3t)$ one obtains $m=t$ and $\langle{T_{\tau_\ast}\rangle} = 2t$, where  $\tau_\ast = t$.
Note also that Eqs. (\ref{our_third_result}) and (\ref{our_first_result}) do not contradict each other since the relation  $h\le2m$ is always valid as shown in \cite{SM}.
Also, Eq. (\ref{our_forth_result}) is in accord with Eq. (\ref{our_first_result}) since $M\le 2m$ for any continuous unimodal probability distribution (see \cite{SM}).


Given this result, it is natural to ask  if the median value can be used to construct the bottom bound of restart performance in the spirit of Eqs. (\ref{our_third_result}) and (\ref{our_forth_result}). 
The answer is no.
A simple counterexample demonstrating that  the inequality $\langle T_{{\cal R}}\rangle\ge C m$, where  $C$ is universal non-zero constant, cannot be valid  is given by the Weibull distribution $P(T)=\frac{k}{\lambda^k}T^{k-1}e^{-(\frac{T}{\lambda})^k}$
 with $0<k<1$ for which $\langle T_{\tau_\ast}\rangle=0$, where  $\tau_\ast\to0$,  and $m> 0$.




\begin{figure}
\centering
\includegraphics[width=\columnwidth]{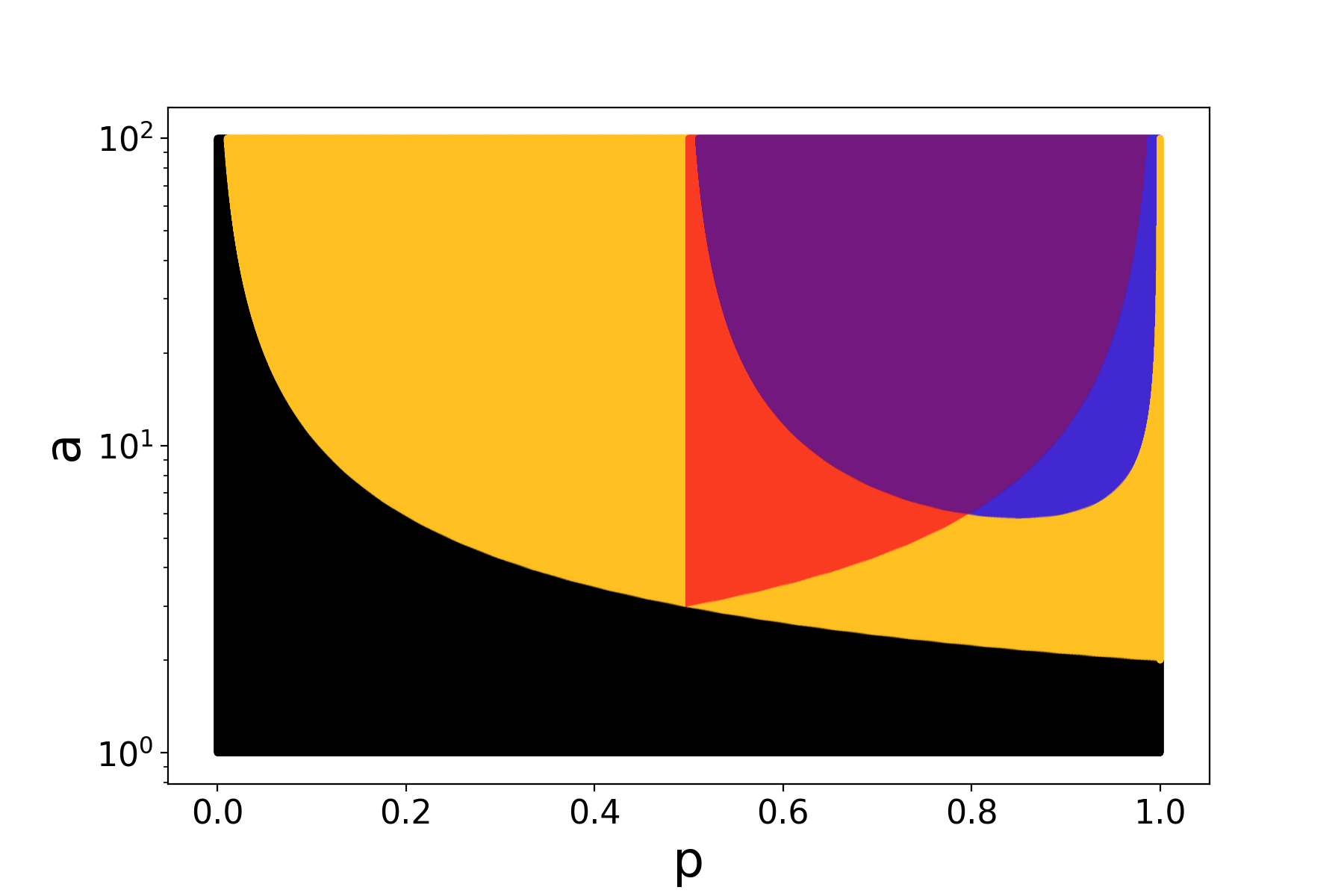}
\label{fig: 5}
\caption{A diagram of restart efficiency for the random process with completion time probability density $P(T) = p\cdot \delta(T-t_1) + (1-p) \cdot \delta(T-t_2)$ in the plane of  dimensionless parameters $p$ and $a=t_2/t_1$. Black area corresponds to the case when restart is not efficient. For the purple area of the diagram restart is efficient and both criteria, $\sigma > \mu$ and $\mu > 2m$, are fulfilled. The region in blue  corresponds to the scenario when restart efficiency is captured only by the inequality $\sigma > \mu$, while  the region in red  -- to the scenario when only the condition $\mu > 2m$ is satisfied. Finally, neither of the two sufficient conditions is met for the orange area of the diagram, but restart is useful nevertheless. }
\end{figure}

\textit{Beyond the mean performance.}
Inequality constraints  given by Eqs. (\ref{our_third_result}), (\ref{our_forth_result}) and  (\ref{our_first_result})  
can be  generalized to higher order statistical moments of random completion time.
First of all, since $\sqrt[k]{\langle T_{{\cal R}}^k\rangle}\ge \langle T_{{\cal R}}\rangle$ for any natural $k$ due to Jensen's inequality \cite{Gallager_2013}, we immediately find from Eq. (\ref{our_third_result}) that 
$\sqrt[k]{\langle T_{{\cal R}}^k\rangle}\ge \frac{1}{4}h$  
for a generic stochastic process under an arbitrary restart protocol.
Likewise,  Eq. (\ref{our_forth_result}) trivially entails inequality
$\sqrt[k]{\langle T_{{\cal R}}^k\rangle}\ge \frac{1}{4}M$ 
which is valid in the case of unimodal completion time distribution.

A similar extension of Eq.(\ref{our_first_result})  is more tricky.
It turns out that the statistical moments of the optimal completion time $T_{\tau_\ast}$  satisfy the inequality
\begin{equation}
\label{our_second_result}
\sqrt[k]{\langle T_{\tau_\ast}^k\rangle} \le 2\sqrt[k]{k!} m.
\end{equation}
To prove Eq. (\ref{our_second_result}) let us assume that the process, which is being restarted periodically in an optimal way, becomes subject to an additional restart protocol ${\cal R}_\Gamma$ characterized by random restart-intervals $\tau_1,\tau_2,\dots$  independently sampled from Gamma distribution $\rho(\tau)=\frac{\beta^k}{\Gamma(k)}\tau^{k-1}e^{-\beta \tau}$ with shape parameter $k$ and infinitesimally small rate parameter $\beta$.
In \cite{SM} we show that this  produces a deferential correction $\langle T_{\tau_\ast+{\cal R}_\Gamma}\rangle-\langle T_{\tau_\ast}\rangle\approx\frac{1}{k!}\left(\langle T_{\tau_\ast}\rangle\langle T^k_{\tau_\ast}\rangle -\frac{1}{k+1}\langle T^{k+1}_{\tau_\ast}\rangle\right)\beta^k$ to the mean completion time  attained by the optimal periodic restart.
Because of the dominance of a periodic restart over other restart strategies, one can be sure that this difference is positive, and therefore
\begin{equation}
\label{constraint}
\frac{\langle T_{\tau_\ast}^k\rangle}{\langle T_{\tau_\ast}\rangle^k}\le k!,
\end{equation}
Together with Eq. (\ref{our_first_result}) this yields Eq. (\ref{our_second_result}).



\textit{Numerical examples.}
For the sake of illustration we explored   several   probability distributions $P(T)$, whose response to restart has been extensively  discussed in the  physical and computer science literature: first-passage time densities for one-dimensional \cite{EM_2011,Evans_review_2020} and  two-dimensional \cite{Evans_2014,Evans_review_2020,Faisant_2021} diffusion  processes, 
first-passage time density for one-dimensional drift-diffusion process \cite{Ray_2019a}, first-passage time density for one-dimensional diffusion in logarithmic potential \cite{Ray_2020},  
log-normal distribution \cite{Schroeder_2001,Reuveni_2014,Lorenz_2018,Lorenz_2021,Eliazar_JPA_2021}, hyper-Erlang distribution \cite{Moorsel_2004,Moorsel_2006,Rotbart_2015,Pal_2019c}, hyper-exponential distribution \cite{Moorsel_2004,Moorsel_2006,Rotbart_2015}, and Pareto distribution \cite{Wu_2006,Lorenz_2018,Lorenz_2021}.
The  numerical parameters associated with the distributions  are given in \cite{SM}.
Numerical data for the average completion time at the optimally chosen frequency of periodic restart are summarized in the diagrams presented in  Fig. \ref{fig:4}.
We see that  all points fall into the  region determined by  Eqs. (\ref{our_third_result}), (\ref{our_forth_result}) and (\ref{our_first_result})  together with the inequalities $h\le2m$ and
$M\le 2m$.


Importantly,  theoretical analyses presented above   does not answer the question of whether the  bounds determined by Eqs. (\ref{our_third_result}) and (\ref{our_forth_result})    are sharp.
Thus, one may expect,  that the  constant $1/4$ entering the right-hand sides of these equations  can potentially be replaced by
one larger.
Numerical data presented in Fig. \ref{fig:4} allow us to argue  that the unknown best possible constants $C_1$ and $C_2$ in the  inequality constraints $\langle T_{{\cal R}}\rangle\le C_1h$ and  $\langle T_{{\cal R}}\rangle\le C_2M$  representing the  exact lower bounds on $\langle T_{{\cal R}}\rangle$  lie in the ranges $1/4\le C_1<5/4$ and $1/4\le C_2<5/3$.


\textit{Corollary 1: Novel criterion  of restart efficiency.}
A notable implication of the upper bound dictated by Eq. (\ref{our_first_result})  is a previously unknown sufficient condition  of when  restart is helpful in  facilitating the process completion.
Namely, as follows from Eq. (\ref{our_first_result}),   inequality $\mu/m> 2$, where $\mu=\langle T\rangle$, guarantees that    there exists finite restart period decreasing the expected completion time.
What is particularly interesting is that this simple inequality makes it possible to capture the benefit of restarting in those cases when  analysis of the relative fluctuation cannot.
In  Fig. 2  we compare the applicability of two  criteria, $\mu/m> 2$ and $\sigma/\mu> 1$, using the mix of two delta-functions  as a model distribution.
Clearly, there is a region of parameters, where the relative fluctuation  is less than unity, $\sigma/\mu< 1$, while $\mu/m> 2$.
More generally, exploiting the well-known probabilistic inequality $|\mu-m|\le \sigma$ \cite{Hotelling_1932,Ocinneide_1990,Mallows_1991}, it is easy to see that the condition $\mu/m> 2$ implies that $\sigma/\mu> 1/2$.
What is more,  the numerical constant $1/2$ cannot be replaced by a larger one: in \cite{SM} we construct an example of probability density $P(T)$ with $\sigma/\mu\to1/2$ and $\mu/m> 2$.




At the same time it should be noted that the opposite scenario, i.e.   $\sigma/\mu> 1$ and $\mu/m< 2$  is also possible (see Fig. 2).
These observations suggest that, since both conditions, $\sigma/\mu>1$ and $\mu/m> 2$,  are sufficient, but by no means necessary, in practice they should be used together to compensate (at least partially) each other's shortcomings.

\textit{Corollary 2: Generalized fluctuation relation for optimal restart.}
Let us also note that  Eq. (\ref{constraint}),  which we have derived for auxiliary purposes, is interesting in itself as it represents a higher-order generalization of the  well-known inequality constraint  $\langle T_{\tau_\ast}^2\rangle/\langle T_{\tau_\ast}\rangle^2\le 2$ first derived by Pal and Reuveni \cite{Reuveni_PRL_2017}. 
Interestingly, for $k=3$,  we find from Eq. (\ref{constraint}) an estimate $\langle T_{\tau_\ast}^3\rangle/\langle T_{\tau_\ast}\rangle^3\le 6$
emphasizing the contrast between best periodic protocol and  optimally tuned Poisson restart for which one  has opposite inequality $\langle T_{r_\ast}^3\rangle/\langle T_{r_\ast}\rangle^3\ge 6$ provided $0<r_\ast<+\infty$, where $r_\ast$ is the optimal restart rate \cite{SM}.


\textit{Conclusion and outlook.}
Revealing explicit performance bounds  is crucial in many areas of science and engineering.
Say, establishment of the Carnot cycle efficiency \cite{Carnot_1824} played a fundamental role for the development of combustion engines and thermal power plants as it sets a bound on the efficiency of any thermodynamic heat engine.
Similarly, Shannon's limit of information capacity \cite{Shannon_1959} has become a guiding principle in the design of communication systems.
Although optimization via restart is widely used in the practice of computer programming and represents active field of academic research in physics, the question of performance limits of this control tool has not been addressed thus far.
In this study, we expressed  these  limits  in terms of simple statistical metrics that can be easily estimated based on finite samples of the process completion time.


Importantly, the presented analysis was grounded on the assumption of instantaneous restart events.
Although this scenario covers overwhelming majority  of  the model problems considered in the literature, it should be kept in mind that in real-life settings restart may be accompanied by some time penalty \cite{Evans_2018b,Ahmad_2019a,Pal_2019a,Bodrova_2020a,Bodrova_2020b,Pal_JPA_2022,Robin_2018a,Reuveni_2014,Reuveni_PRL_2016,Friedman_2020a}.
Say, in the context of single molecule enzyme kinetics, where restart occurs naturally by virtue of intermediate  dissociation,   some time is required to enzyme which unbinds from its substrate to find a new one in surrounding solution \cite{Reuveni_2014,Reuveni_PRL_2016,Robin_2018a}.
Similarly,  restart of computer program typically involves a time overhead.
Also, models with noninstantaneous restarts provide more realistic pictures of colloidal particle diffusion with resetting \cite{Friedman_2020a}.
How does accounting for delays modifies the bounds constructed here?
The straightforward generalization of arguments leading to Eq. (\ref{our_first_result}) brings us to the following simple result (see \cite{SM}):   $\langle T_{\tau_\ast}\rangle \le 2m+2\langle T_{on}\rangle$,
where $\langle T_{on}\rangle$ is the expectation of generally distributed time penalty which
collectively accounts for any delays that may arise prior to completion attempt.
A similar generalization of the lower bounds given by Eqs. (\ref{our_third_result}) and (\ref{our_forth_result}) is an important (and apparently sophisticated)  task for future research.





\acknowledgments

The work was supported by the Russian Science Foundation,
project no. 22-72-10052.

\begin{widetext}

\vspace{0.5cm}

\qquad\qquad\qquad\qquad\qquad\qquad\qquad\qquad \Large \underline{Supplementary Material}

\normalsize

\section{I. Derivation of Eq. (1) in the main text}

Let us consider a more general situation than the one described in the main text.
Namely, we assume that process initiation and each restart event entail  time delays representing statistically independent realizations of some generally distributed random variable  $T_{on}$.
Then, the random completion  time $T_\tau$ in the presence of regular restart with period  $\tau$ obeys the stochastic renewal equation \cite{Reuveni_PRL_2017}
\begin{eqnarray}
\label{chain1}
&&T_\tau=T_{on}+TI(T< \tau)+(\tau+T_\tau') I(T\ge \tau),
\end{eqnarray}
where $T_\tau'$ is a statistically independent copy of $T_\tau$, and $I(...)$ denotes the indicator random variable which is equal to unity when the inequality in its argument is justified and is zero otherwise.
Performing averaging over the statistics of $T$ and $T_{on}$ we obtain
\begin{eqnarray}
\langle T_\tau\rangle=\langle T_{on} \rangle+ \langle TI(T< \tau)\rangle+      \tau \langle I(T\ge \tau)\rangle +\langle T_\tau' I(T\ge \tau)\rangle.
\end{eqnarray}
Since random variables $T$, $T_\tau'$ and $T_{on}$ are statistically independent we can rewritte
\begin{eqnarray}
\langle T_\tau\rangle=\langle T_{on} \rangle+\langle TI(T< \tau)\rangle+      \tau \langle I(T\ge  \tau)\rangle + \langle T_\tau' \rangle \langle  I(T\ge \tau)\rangle.
\end{eqnarray}
Next,  taking into account that $\langle T_\tau' \rangle=\langle T_\tau \rangle$ one obtains
\begin{equation}
\label{mean_T_tau_1}
\langle T_{\tau}\rangle= \frac{\langle T_{on} \rangle +\langle TI(T< \tau)\rangle+\tau\langle I(T\ge  \tau)\rangle }{ \langle I(T< \tau)\rangle}.
\end{equation}
If the original stochastic process halts with unit probability, then the random variable $T$ is described by probability density $P(T)$ with standard normalization  $\int_0^\infty dTP(T)=1$ and we find  
$\langle TI(T< \tau)\rangle=\int_0^\tau P(T)TdT$,  $\langle I(T\ge \tau)\rangle=\int_\tau^\infty P(T)dT$, and  $1-\langle I(T\ge \tau)\rangle=\int_0^{\tau} P(T)dT$.
Substituting these expressions into Eq. (\ref{mean_T_tau_1}) gives 
\begin{equation}
\label{completion_time11}
\langle T_{\tau}\rangle = \frac{\langle T_{on} \rangle+\int_0^{\tau} P(T)TdT+\tau\int_{\tau}^\infty P(T)dT }{\int_0^{\tau} P(T)dT}.
\end{equation}
At $T_{on}=0$ this result coincides with Eq. (1) in the main text.

%
%
%
%

\section{II. Simple proof of global dominance of strictly regular restart}

Luby et al. \cite{Luby_1993} showed that strictly periodic restart strategy is universally optimal for discrete completion time distributions.
Pal and Reuveni \cite{Reuveni_PRL_2017} proved  dominance of strictly periodic restart over any stochastic restart protocol with independent and identically distributed inter-restarts intervals in a continuous setting.
Finally,  the global dominance of strictly periodic restart  in the space of all possible restart protocols for continuous case follows from Lemma 3 in Lorenz's paper \cite{Lorenz_2021}.
In order to make the presentation of our results complete and intuitively clear, here we developed a more simple line of arguments in support of  the optimality of periodic restart.
Noteworthy, our proof is not only simpler, but also more general than the previous ones, since we take into account a non-zero delay $T_{on}$ before starting/restarting.

Let $\tau_{\ast}$ be the best period of regular restart protocol for a given stochastic process, i.e. 
\begin{equation}
\label{app0}
\langle T_{\tau_{\ast}}\rangle\le \langle T_{\tau}\rangle
\end{equation}
 for any $\tau\ge 0$.
Also, let ${\cal R}_\ast=\{\tau_1^\ast,\tau_2^\ast,...\}$ be an optimal restart protocol for the same process.
By the definition of optimal protocol this means that   
\begin{equation}
\label{app1}
\langle T_{{\cal R}_\ast}\rangle\le \langle T_{{\cal R}}\rangle
\end{equation}
for any ${\cal R}$.
Below we show that $\langle T_{\tau_{\ast}}\rangle= \langle T_{{\cal R}_\ast}\rangle$.

The random completion time in the presence of restart events scheduled accordingly to the protocol ${\cal R}_\ast$ can be represented as 
\begin{equation}
\label{app2}
T_{{\cal R}_\ast}=T_{on}+TI(T<\tau_1^\ast)+(\tau_1^\ast+T_{{\cal R}_\ast'})I(T\ge \tau_1^\ast)
\end{equation}
where ${\cal R}_\ast'=\{\tau_2^\ast,\tau_3^\ast,...\}$.
Averaging over the statistics of original process yields 
\begin{equation}
\label{app3}
\langle T_{{\cal R}_\ast}\rangle=\langle T_{on}\rangle+\langle TI(T<\tau_1^\ast)\rangle+\tau_1^\ast\langle I(T\ge\tau_1^\ast)\rangle +\langle T_{{\cal R}_\ast'}\rangle\langle I(T\ge\tau_1^\ast)\rangle,
\end{equation}
where we exploited statistical independence of $T$ and $T_{{\cal R}_\ast'}$.
Since  ${\cal R}_\ast$ is optimal, then $\langle T_{{\cal R}_\ast'}\rangle\ge \langle T_{{\cal R}_\ast}\rangle$, and, therefore,  
\begin{equation}
\label{app4}
\langle T_{{\cal R}_\ast}\rangle\ge \langle T_{on}\rangle+\langle TI(T<\tau_1^\ast)\rangle+\tau_1^\ast\langle I(T\ge\tau_1^\ast)\rangle +\langle T_{{\cal R}_\ast}\rangle\langle I(T\ge\tau_1^\ast)\rangle,
\end{equation}
and
\begin{equation}
\label{app5}
\langle T_{{\cal R}_\ast}\rangle\ge \frac{\langle T_{on} \rangle +\langle TI(T< \tau_1^\ast)\rangle+\tau_1^\ast\langle I(T\ge  \tau_1^\ast)\rangle }{ \langle I(T< \tau_1^\ast)\rangle}
\end{equation}
Comparing right-hand sides of  Eq. (\ref{app5}) and  Eq. (\ref{mean_T_tau_1}),  one obtains
\begin{equation}
\label{app6}
\langle T_{{\cal R}_\ast}\rangle\ge \langle T_{\tau_1^\ast}\rangle.
\end{equation}
From Eqs. (\ref{app0}), (\ref{app1}) and (\ref{app6}) we may conclude that $\langle T_{{\cal R}_\ast}\rangle=\langle T_{\tau_\ast}\rangle$, and, therefore, $\langle T_{\tau_\ast}\rangle\le \langle T_{{\cal R}}\rangle$ for any ${\cal R}$.

\section{III. Derivation of Eq. (2) in main text}

Luby et al. \cite{Luby_1993}  stated Eq. (2) from the main  text for discrete probability distributions. More recently, Lorenz \cite{Lorenz_2021} rigorously showed this property for continuous distributions.  
Here we briefly remind the arguments presented in \cite{Lorenz_2021}.
Note that in this part of analysis we adopt $T_{on}=0$.



If  $\langle T_{\tau_\ast} \rangle\ge \frac{\tau_\ast}{2Pr[T\le \tau_\ast]}$, then 
the inequality $\langle T_{\tau_\ast} \rangle\ge \frac{1}{4} \min_{\tau} \frac{\tau}{Pr[T\le \tau]}$ is evident. 

Next, we  assume that $\langle T_{\tau_\ast} \rangle<\frac{\tau_\ast}{2Pr[T\le \tau_\ast]}$.
Let us introduce a variable   $\tilde{T}\equiv \min(T,\tau_\ast)$.
It follows from Eq. (\ref{mean_T_tau_1}) that   $\langle T_{\tau}\rangle =\frac{\langle \min(T,\tau)\rangle}{Pr[T\le \tau_\ast]}$, and, therefore,   $\langle\tilde{T} \rangle=\langle T_{\tau_\ast}\rangle Pr[T\le \tau_\ast] $.
Next, due to the assumption $\tau_\ast >2\langle T_{\tau_\ast} \rangle Pr[T\le \tau_\ast]$ and the  Markov's inequality  $Pr[\tilde{T} > 2\langle\tilde{T} \rangle]\le \frac12$ we find
$
Pr[T>2\langle T_{\tau_\ast} \rangle Pr[T\le \tau_\ast]]=Pr[\tilde{T}>2\langle T_{\tau_\ast} \rangle Pr[T\le \tau_\ast]]=Pr[\tilde{T}>2\langle\tilde{T} \rangle]\le \frac12$. 
Therefore
\begin{equation}
\label{proof1}
Pr[T\le 2\langle T_{\tau_\ast} \rangle Pr[T\le \tau_\ast]]\ge\frac12.
\end{equation}
Now, let us denote $t\equiv 2\langle T_{\tau_\ast} \rangle Pr[T\le \tau_\ast]$.
It follows from the inequalities (\ref{proof1}) and $Pr[T\le \tau]\le 1$
 that 
\begin{eqnarray}
\min_{\tau} \frac{\tau}{Pr[T\le \tau]}\le \frac{t}{Pr[T\le t]}=\frac{2\langle T_{\tau_\ast} \rangle Pr[T\le \tau_\ast]}{Pr[T\le 2\langle T_{\tau_\ast} \rangle Pr[T\le \tau_\ast]]}\le 4 \langle T_{\tau_\ast} \rangle. 
\end{eqnarray}
This completes the proof.

\section{IV. Checking  consistency of  Eqs. (3), (4) and (5) in main text}

Equations (3) and (5) do not contradict each other since $h\le2m$ for any probability density $P(T)$.
Indeed, applying  Markov's inequality we find  $1/2=\text{Pr}[T\le m]=\text{Pr}[\omega\ge \frac{1}{m}]\le \langle \omega\rangle m=T_{1/2}/h$, where $\omega=1/T$.

Also, Eq. (4) does not contradict Eq. (5) since $M\le 2m$ for any continuous unimodal probability density $P(T)$.
To prove this, assume the contrary, that is, let $M> 2m$ be true for some non-negative random variable $T$.
By the virtue of definition we have  $\int_{0}^{m}P(T)dT=\int_{m}^{+\infty}P(T)dT=1/2$.
Further, it follows from the definition of the mode  that the function $P(T)$ is non-decreasing on the interval $[0,M]$.
Then, on the one hand $\int_{m}^{M}P(T)dT> \int_{m}^{2m}P(T)dT\ge \int_{0}^{m}P(T)dT=1/2$, and on the other $\int_{m}^{M}P(T)dT\le \int_{m}^{+\infty}P(T)dT=1/2$.
We got a contradiction.

One may expect that this is not the first time the simple  relations  $h\le2m$  and $M\le 2m$ have been proven, but we have not found an appropriate reference.

\section{V. Derivation of Eq. (6) in main text}

Assume that a stochastic process with finite mean  completion time $\langle T\rangle$ becomes subject to the stochastic restart protocol ${\cal R}_{\Gamma}=\tau_1,\tau_2,\dots$, where  random intervals between instantaneous restarts are independently sampled from the  Gamma distribution   $\rho(\tau)=\frac{\beta^k}{\Gamma(k)}\tau^{k-1}e^{-\beta \tau}$ with differentially small rate parameter  $\beta$.
Let us investigate how this modifies the average completion time. 

In the presence of the stochastic restart, the random completion time $T_{{\cal R}_\Gamma}$ obeys the following renewal equation
\begin{eqnarray}
\label{chain11111}
&&T_{{\cal R}_\Gamma}=TI(T< \tau)+(\tau+T_{{\cal R}_\Gamma}') I(T\ge \tau),
\end{eqnarray}
where $T_{{\cal R}_\Gamma}'$ is a statistically independent replica of $T_{{\cal R}_\Gamma}$.
Let us average this relation over the statistics of original process and of the inter-restart intervals.
This gives
\begin{equation}
\langle T_{{\cal R}_\Gamma}\rangle=\langle TI(T< \tau)\rangle+ \langle\tau I(T\ge  \tau)\rangle+
\langle T_{{\cal R}_\Gamma}'  I(T\ge \tau)\rangle.
\end{equation}
Since $\langle TI(T< \tau)\rangle=\int_0^\infty dT P(T)T  \int_T^\infty d\tau \rho(\tau) $, $ \langle\tau I(T\ge  \tau)\rangle=\int_0^\infty dT P(T)  \int_0^T d\tau \rho(\tau)\tau$, $\langle T_{{\cal R}_\Gamma}'  I(T\ge \tau)\rangle=\langle T_{{\cal R}_\Gamma}' \rangle\langle I(T\ge  \tau)\rangle$, $1-\langle I(T\ge \tau)\rangle=\int_0^\infty dT P(T)  \int_T^\infty d\tau \rho(\tau)$ and $\langle T_{{\cal R}_\Gamma}' \rangle=\langle T_{{\cal R}_\Gamma}\rangle$, we find the closed-form expression for the expected completion time 
\begin{equation}
\label{stochastic}
\langle T_{{\cal R}_\Gamma}\rangle=\frac{\int_0^\infty dT P(T)T  \int_T^\infty d\tau \rho(\tau) +\int_0^\infty dT P(T)  \int_0^T d\tau \rho(\tau)\tau}{\int_0^\infty dT P(T)  \int_T^\infty d\tau \rho(\tau)}.
\end{equation}
In leading order in the small parameter $\beta$ we obtain
\begin{eqnarray}
&&\int_0^\infty dT P(T)T  \int_T^\infty d\tau \rho(\tau)=\langle T\rangle -\int_0^\infty dT P(T)T  \int_0^T d\tau \rho(\tau)\approx\\
&&\approx  \langle T\rangle -\frac{\beta^k}{(k-1)!}\int_0^\infty dT P(T)T  \int_0^T d\tau \tau^{k-1}=  \langle T\rangle-\frac{\beta^k}{k!}\langle T^{k+1}\rangle,\\
&&\int_0^\infty dT P(T)  \int_0^T d\tau \rho(\tau)\tau\approx  \frac{\beta^k}{(k-1)!}\int_0^\infty dT P(T)  \int_0^T d\tau \tau^k=\frac{k\beta^k}{(k+1)!}\langle T^{k+1}\rangle,\\
&&\int_0^\infty dT P(T)  \int_T^\infty d\tau \rho(\tau)=1 -\int_0^\infty dT P(T)  \int_0^T d\tau \rho(\tau)\approx\\
&& \approx 1 - \frac{\beta^k}{(k-1)!}\int_0^\infty dT P(T)  \int_0^T d\tau \tau^{k-1} = 1-\frac{\beta^k}{k!}\langle T^k\rangle,
\end{eqnarray}
and, therefore,
\begin{equation}
\label{app10}
\langle T_{{\cal R}_\Gamma}\rangle\approx \langle T\rangle +\frac{1}{k!}\left(\langle T\rangle\langle T^k\rangle -\frac{1}{k+1}\langle T^{k+1}\rangle\right)\beta^k.
\end{equation}

Now suppose that the stochastic restart protocol is applied to a process that is already subject to optimal periodic restart.
Equation (\ref{app10}) tells us that the resulting mean completion time is given by $\langle T_{\tau_\ast+{\cal R}_{\Gamma}}\rangle=\langle T_{\tau_\ast}\rangle+\frac{1}{k!}\left(\langle T_{\tau_\ast}\rangle\langle T^k_{\tau_\ast}\rangle -\frac{1}{k+1}\langle T^{k+1}_{\tau_\ast}\rangle\right)\beta^k$.
Because of the dominance of a periodic restart over other restart strategies, one can be sure that $\langle T_{\tau_\ast+{\cal R}_{\Gamma}}\rangle \ge \langle T_{t_\ast}\rangle$, and therefore  $\langle T^{k+1}_{\tau_\ast}\rangle\le (k+1) \langle T_{\tau_\ast}\rangle\langle T^k_{\tau_\ast}\rangle$ for any natural $k$.
This immediately implies Eq. (6). 

\section{VI. Example of distribution with $\sigma/\mu = 1/2$ and $\mu/m > 2$}

Consider a distribution of the form
\begin{equation}
    P(T) =  \left(\frac{1}{2} + \varepsilon\right) \cdot \delta(T-t_1) + \left( \frac{1}{2} - \varepsilon \right) \cdot \delta(T-(3 + 6\varepsilon)\cdot t_1),  
\end{equation}
where $0 < \varepsilon \ll 1$ and $t_1>0$. Its median completion time $m$ is equal to $t_1$, and its mean is given by
\begin{equation}
    \mu\equiv\langle T \rangle = \left(\frac{1}{2} + \varepsilon\right) \cdot t_1 + \left( \frac{1}{2} - \varepsilon \right) \cdot t_1\cdot(3 + 6\varepsilon) = t_1 \cdot \left(2 + \varepsilon  - 6\varepsilon^2\right) > 2 m,
\end{equation}
where the last inequality is fulfilled due to positivity and infinitesimality of $\varepsilon$. Ratio of standard deviation to the mean is thus given by
\begin{equation}
    \frac{\sigma}{\mu} = \frac{(1+3\varepsilon)\cdot \sqrt{1-4\varepsilon^2}}{2+\varepsilon-6\varepsilon^2} = \frac{1+3\varepsilon}{2} \cdot \left(1 - 2\varepsilon^2 + O(\varepsilon^4)\right)\cdot \left(1 - \frac{\varepsilon}{2} + \frac{13}{4}\varepsilon^2 + O(\varepsilon^3)\right) = \frac{1}{2} + \frac{5\varepsilon}{4} + O(\varepsilon^2) \rightarrow \frac{1}{2}.
\end{equation}

\section{VII. Details of distributions used in numerical simulations}

Probability distributions, considered in Fig. 1 in main text, and their corresponding numerical parameters are as follows.  
In all investigated cases restart is beneficial, i.e. there exists $\tau_\ast$ such that $\langle T_{\tau_\ast}\rangle<\langle T\rangle$.

1)First-passage-time distribution for 1d-diffusion process (L{\' e}vi-Smirnov distribution) \cite{EM_2011}:
\begin{equation}
P(T)=\frac{L}{2\sqrt{\pi D}T^{3/2}}\exp\left(-\frac{L^2}{4DT}\right),
\end{equation}
calculated for $L^2/D = 1$, where $D$ is the diffusion coefficient, and $L$ is the initial distance to the target. 

2) First-passage-time distribution for 2d-diffusion process \cite{Faisant_2021}:
\begin{equation}
 P(T) = \frac{2}{\pi}\int_0^{\infty}\frac{dx}{R^2} e^{-DTx^2/R^2} \cdot x \cdot \frac{Y_0(\frac{x\cdot r}{R}) \cdot J_0(x)-J_0(\frac{x\cdot r}{R}) \cdot Y_0(x)}{J_0^2(x)+Y_0^2(x)},
\end{equation}
where $J_0(x)$ and $Y_0(x)$ are the Bessel functions of the first and the second kind respectively. Calculated for $D = 1$ and $R/r = 0.4$, where $D$ is the diffusion coefficient, $r$ - is the radius of absorbing disk and $R$ - distance between starting position of the particle and centre of the disk.

3) First-passage time distribution for 1d-diffusion in logarithmic potential \cite{Ray_2020}:
\begin{equation}
P(T) =\frac{T^{-(v+1)}}{\Gamma(v)}\left(\frac{x_0^2}{4 D}\right)^v \exp \left(-\frac{x_0^2}{4 D T}\right),
\end{equation}
calculated for $\nu = 3$, $x_0 = 1$, $D = 1$.

4) First-passage time distribution for drift-diffusion process to an absorbing boundary \cite{Ray_2019a}:
\begin{equation}
P(T) =\frac{L}{\sqrt{4 \pi D T^3}} \exp \left[-\frac{(L-V T)^2}{4 D T}\right]
\end{equation}
calculated for $L = 2$, $V = 0.5$, $D = 1$.


5) Mix of two exponential distributions (double exponential distribution): 
\begin{equation}
P(T)=p k_1 e^{-k_1 T}+(1-p) k_2 e^{-k_2 T},
\end{equation}
calculated for ($k_1 = 0.5$, $k_2 = 2$, $p = 1/3$), and ($k_1 = 1$, $k_2 = 10.25$, $p = 0.1025$).

6) Mix of three exponential distributions (triple exponential distribution):
\begin{equation}
P(T)=p_1k_1 e^{-k_1 T}+p_2k_2 e^{-k_2 T} + (1-p_1-p_2)k_3e^{-k3 T},
\end{equation}

calculated for $k_1 = 1$, $p_1 = 0.5$, $k_2 = 6$, $p_2 = 0.2$, $k_3 = 15$.

7) Mix of two Erlang distributions (double Erlang distribution):
\begin{equation}
P(T)=p k_1^2 T e^{-k_1 T}+(1-p) k_2^2 T e^{-k_2 T},
\end{equation}
calculated for ($k_1 = 22.88$, $k_2 = 1.42$, $p = 0.31$), and ($k_1 = 1.42$, $k_2 = 22.88$, $p = 0.31$),  

8) Log-Normal distribution:
\begin{equation}
P(T) = \frac{1}{T \sigma \sqrt{2 \pi}} \exp \left(-\frac{(\ln (T)-\mu)^{2}}{2 \sigma^{2}}\right),
\end{equation}
calculated for ($\mu = 0$, $\sigma = 1$) and ($\mu = 0$, $\sigma = 0.75$).




9) Pareto distribution:
\begin{equation}
P(T)=\left\{\begin{array}{ll}
\alpha k^\alpha T^{-1-\alpha} & T \geq k, \\
0 & T<k,
\end{array}\right.
\end{equation}
calculated for ($k = 1$, $\alpha = 1$) and ($k = 1$, $\alpha = 3$).

10) Sum of two delta functions:
\begin{equation}
    P(T) = \frac{1}{2} \delta(T - t) + \frac{1}{2} \delta(T - 3t).
\end{equation}



\section{VIII. Generalization of Eq. (5) to the case of non-instantaneous restarts and non-zero probability of non-stopping}

Here we generalize Eq. (5) from  the main text  assuming  that (a)  the original stochastic process ends for finite time with probability $0<q\le 1$, and (b) any start/restart event is accompanied by a random time penalty $T_{on}$. 

Obviously,   Eq. (\ref{mean_T_tau_1}) can be rewritten  as
\begin{equation}
\langle T_{\tau}\rangle=\frac{\langle T_{on} \rangle +min(T,\tau) }{Pr[T< \tau]},
\end{equation}
where $min(T,\tau)$  is the
minimum of $T$ and $\tau$ and we used the identity $\langle I(T< \tau)\rangle=Pr[T< \tau]$. 
Next, since $\langle T_{\tau_\ast}\rangle\le \langle T_{\tau}\rangle$  and $min(T,\tau)\le \tau$, one immediately obtains an estimate 
\begin{equation}
\label{completion_time1115}
\langle T_{\tau_\ast}\rangle \le \frac{\tau+ \langle T_{on} \rangle}{Pr[T< \tau]}.
\end{equation}
Let us denote as $m_s$ the median completion time of the halting realizations.
Substituting  $m_s$ for  $\tau$ in the last inequality yields
\begin{equation}
\label{completion_time1116}
\langle T_{\tau_\ast}\rangle \le \frac{2(m_s+ \langle T_{on} \rangle)}{q},
\end{equation}
where we took into account that $Pr[T< m_s]=\frac{1}{2}\times q$.

At $q=1$ and $T_{on}=0$,  Eq. (\ref{completion_time1116}) reduces to Eq. (5) in the main text.
The special cases $q<1$, $T_{on}=0$ and $q=1$, $T_{on}>0$ enter the main text as the inline formulas.

%

\section{IX. Derivation of inequality $\langle T_{r_\ast}^3\rangle\ge 6 \langle T_{r_\ast}\rangle^3$ for optimally tuned Poisson restart}

The mean completion time of the process subject to Poisson restart at rate $r$ is equal to  (see \cite{Reuveni_PRL_2016})
\begin{equation}\label{eq:T_r_mean_main}
\langle T_r\rangle=\frac{1-\tilde{P}(r)}{r\tilde{P}(r)},
\end{equation}
where $\tilde{P}(r)=\int_0^\infty dT P(T)e^{-rT}$ -- is the Laplace transform of the probability density $P(T)$ .

Let $r_\ast$ be the optimal restart rate bringing $\langle T_r\rangle$ to a minimum so that 
\begin{eqnarray}
\label{extr1}
\frac{d\langle T_{r_\ast}\rangle}{dr}=0,
\end{eqnarray}
and
\begin{eqnarray}
\label{extr2}
\frac{d^2\langle T_{r_\ast}\rangle}{dr^2}\ge 0.
\end{eqnarray}

Let us assume that the process, which is being restarted at rate $r_\ast$, becomes subject to additional stochastic (Poisson) restart events with infinitesimally small rate $\delta r$.  
Due to the additive property of Poisson events the resulting performance is given by 
\begin{equation}
\label{eq:T_r_mean_main02}
\langle T_{r_\ast+\delta r}\rangle=\langle T_{r_\ast}\rangle+\delta r\frac{d\langle T_{r_\ast}\rangle}{dr}+\delta r^2\frac{d^2\langle T_{r_\ast}\rangle}{dr^2}+o(\delta r^2).
\end{equation}
At the same time, using Eq. (\ref{eq:T_r_mean_main}) we obtain
\begin{eqnarray}
&&\langle T_{r_\ast+\delta r}\rangle=\frac{1-\tilde{P}_{r_\ast}(\delta r)}{\delta r\tilde{P}_{r_\ast}(\delta r)}=\frac{1-\int_0^\infty  e^{-\delta r T_{r_\ast}}P_{r_\ast}(T_{r_\ast})dT_{r_\ast}}{\delta r\int_0^\infty  e^{-\delta r T_{r_\ast}}P_{r_\ast}(T_{r_\ast})dT_{r_\ast}}=\\
\label{eq:T_r_mean_main03}
&&=\langle T_{r_\ast}\rangle+\delta r \left(\langle T_{r_\ast}\rangle^2-\frac12 \langle T_{r_\ast}^2\rangle\right)+\delta r^2\left(\frac16\langle T_{r_\ast}^3\rangle-\langle T_{r_\ast}^2\rangle\langle T_{r_\ast}\rangle+\langle T_{r_\ast}\rangle^3 \right)+o(\delta r^2),
\end{eqnarray}
where $P_{r_\ast}(T_{r_\ast})$ denotes the probability density of the random completion time under optimal Poisson restart.
and  $\tilde{P}_{r_\ast}(r)=\int_0^\infty  e^{-r T_{r_\ast}}P_{r_\ast}(T_{r_\ast})dT_{r_\ast}$ is its Laplace transform.

Comparing Eqs. (\ref{eq:T_r_mean_main02}) and (\ref{eq:T_r_mean_main03}) one obtains $\frac{d\langle T_{r_\ast}\rangle}{dr}=\langle T_{r_\ast}\rangle^2-\frac12 \langle T_{r_\ast}^2\rangle$ and $\frac{d^2\langle T_{r_\ast}\rangle}{dr^2}=\frac16\langle T_{r_\ast}^3\rangle-\langle T_{r_\ast}^2\rangle\langle T_{r_\ast}\rangle+\langle T_{r_\ast}\rangle^3$. 
Substituting these relations into  Eqs. (\ref{extr1}) and (\ref{extr2}) yields
\begin{equation}
\langle T_{r_\ast}^3\rangle\ge 6 \langle T_{r_\ast}\rangle^3.
\end{equation} 


\end{widetext}

{}


\begin{thebibliography}{99}
\bibitem{note0} 
Noteworthy, once the probability $q=Pr[T<+\infty]$ of halting for a given stochastic process is known, one can improve the estimate given by Eqs. (\ref{our_third_result})
 as follows: $\langle T_{\cal R}\rangle\ge \frac14 h/q$.
\end{thebibliography}
\end{document}